\documentclass{article}  
\newcommand{\refcite}[1]{\cite{#1}}  
\newcommand{\catchline}{}  
\renewcommand{\author}[1]{#1} 
\newcommand{\address}[1]{#1}  
\newcommand{\keywords}[1]{Keywords: #1}  

\usepackage{moreverb}
\usepackage{xspace}
\usepackage{epsfig}

\newcommand{\Gfour}{{\sc Geant4}\xspace}

\newcommand{\GVS}{\Gfour Visualisation System\xspace}
\newcommand{\GVM}{\Gfour Visualisation Manager\xspace}

\newcommand{\vs}{visualisation system\xspace}
\newcommand{\vm}{visualisation manager\xspace}

\begin{document}

\hfill MAN/HEP/2012/19

\markboth{Allison et al.}{The \Gfour Visualisation System}

\catchline{}{}{}{}{}


\vspace{1cm}
\begin{center}
\LARGE The \Gfour Visualisation System---a multi-driver graphics system
\end{center}
\vspace{1cm}


\author{John Allison} 
\address{School of Physics and Astronomy, The University of Manchester, Manchester, M13 9PL, UK\\
and \Gfour Associates International Ltd., 9 Royd Terrace, Hebden Bridge, HX7 7BT, UK\\
John.Allison@g4ai.org}

\author{Laurent Garnier}
\address{Laboratoire de l'Acc\'{e}l\'{e}rateur Lin\'{e}aire, 91898 Orsay cedex, France\\
garnier@lal.in2p3.fr}

\author{Akinori Kimura}
\address{Ashikaga Institute of Technology, 268-1 Omae-cho, Ashikaga City, Tochigi, 326-8558 Japan\\
akimura@ashitech.ac.jp}

\author{Joseph Perl}
\address{SLAC National Accelerator Laboratory, 2575 Sand Hill Road, Menlo Park, CA 94025-7015, USA\\
perl@slac.stanford.edu}



\vspace{12pt}
\begin{center}
Submitted to the International Journal of Modeling, Simulation, and\\
Scientific Computing.
\end{center}

\begin{abstract}
From the beginning the \Gfour Visualisation System was designed to support several simultaneous graphics systems written to common abstract interfaces. Today it has matured into a powerful diagnostic and presentational tool. It comes with a library of models that may be added to the current scene and which include the representation of the \Gfour geometry hierarchy, simulated trajectories and user-written hits and digitisations. The workhorse is the OpenGL suite of drivers for X, Xm, Qt and Win32. There is an Open Inventor driver. Scenes can be exported in special graphics formats for offline viewing in the DAWN, VRML, HepRApp and gMocren browsers. PostScript can be generated through OpenGL, Open Inventor, DAWN and HepRApp. \Gfour's own tracking algorithms are used by the Ray Tracer. Not all drivers support all features but all drivers bring added functionality of some sort. This paper describes the interfaces and details the individual drivers.
\end{abstract}


\section{Introduction}	

\Gfour arose out of a desire to take advantage of a new programming paradigm, namely object-oriented programming, to create a toolkit for the simulation of the passage of particles and radiation though matter that could be developed and maintained by physicists expert in radiation and particle interactions from around the world. The initiative for a new toolkit came from KEK and CERN in the early 1990s and this led to a research and development project, RD44, funded by CERN from 1994-1998 in the run up to the commissioning of the LHC, in which the paradigm was realised in the C++ programming language. Many other institutes, notably SLAC, also provided support. Many physicists from around the world gave their time and effort as part of their research programmes. \Gfour was born out of this project in 1999 as an independent collaboration with its own collaboration agreement, providing open code that has found applications in medicine, space and related domains as well as nuclear and particle physics that were its origins. The modular design, which is a natural consequence of object-oriented programming, makes possible the development by many authors in parallel in a well defined and safe manner, a process that continues to the present day. A more detailed history of \Gfour can be found in the original general paper.\cite{Geant4} Some more recent developments and applications are described in Reference~\refcite{DevsAndApps}.

The visualisation system was designed to facilitate the development of \Gfour and its applications. It was written to a basic abstract interface and itself introduced interfaces for multiple drivers. This was first described in Reference~\refcite{VisGen} and implementation details are given in the Toolkit Developers Guide.\cite{UGT}  The current paper brings this up to date and describes progress since then.

A good practical introduction to the \GVS is given in Reference~\refcite{SLACPresentation}.

\section{Overview of the \GVS}

Every application that needs to use the \GVS must instantiate a {\tt G4VisManager} and register visualisation drivers as desired from those available on the computing platform. (This is done automatically should the application be built through the \Gfour build system and {\tt G4VisExecutive} as described below.)  The application communicates with the \vm through basic abstract interfaces, also described below.

The \GVM has the ability to build ``scenes'' ({\tt G4Scene} objects) with any number of geometrical (detector) components, axes, annotations and (at the end of event or run) the potential to draw particle trajectories, hits (representations of effects, for example, energy deposit, of particles) and digitisations (ultimate signals from sensitive components---``digits'' for short). It defines,  through the {\tt G4VSceneHandler} interface, ``scene handlers'' that translate the scene into messages for a particular graphics system and, through the {\tt G4VViewer} interface, ``viewers'' that render to the final device (screen, file or terminal). Any scene can be associated with any scene handler and viewed with any viewer. A scene handler-viewer(s) combination is referred to as a driver. One may instantiate any number of drivers of any type, each with any number of viewers, and switch between them.

It is a natural consequence of object oriented design that any driver that conforms to these interfaces can be accessed by the \vm and we take advantage of this to write multiple drivers with different characteristics and qualities---for example, OpenGL and Open Inventor for fast drawing (with varying degrees of interactivity according to the computing platform), DAWN for high quality PostScript output, HepRepFile for scene browsing, VRML for virtual reality, RayTracer for photorealistic images, gMocren for medical images and ASCIITree for a text dump.

The \GVM keeps a list of scenes, scene handlers and viewers.  There is always a current viewer serviced by its scene handler with its scene. Drawing requests made by the application or re-issued by the \vm are always made to the current viewer.  If a scene is changed, or on user request, all views of that scene are rebuilt.

The \GVS is implemented as a ``plug-in''.  It may use any part of the toolkit but itself may only be used by a \Gfour user application or by the toolkit itself through the basic abstract interfaces in a protected way, as described below.

The \GVS is currently being used across a wide range of \Gfour applications in high energy physics, nuclear physics, space and medicine.

\section{The Basic Abstract Interfaces}
\label{secAbstractInterface}

Following our object-oriented design, any Geant4 Visualisation Manager must implement the G4VVisManager interface.  This is the working interface for all drawing messages. As one can see from Figure \ref{figG4VVisManager}, it is generously endowed with possibilities.

\begin{figure}
\begin{center}
{\scriptsize
\begin{boxedverbatim}
class G4VVisManager {
public:
  static G4VVisManager* GetConcreteInstance ();
  virtual void Draw (const G4Circle&,
    const G4Transform3D& objectTransformation = G4Transform3D()) = 0;
  virtual void Draw (const G4NURBS&,
    const G4Transform3D& objectTransformation = G4Transform3D()) = 0;
  virtual void Draw (const G4Polyhedron&,
    const G4Transform3D& objectTransformation = G4Transform3D()) = 0;
  virtual void Draw (const G4Polyline&,
    const G4Transform3D& objectTransformation = G4Transform3D()) = 0;
  virtual void Draw (const G4Polymarker&,
    const G4Transform3D& objectTransformation = G4Transform3D()) = 0;
  virtual void Draw (const G4Scale&,
    const G4Transform3D& objectTransformation = G4Transform3D()) = 0;
  virtual void Draw (const G4Square&,
    const G4Transform3D& objectTransformation = G4Transform3D()) = 0;
  virtual void Draw (const G4Text&,
    const G4Transform3D& objectTransformation = G4Transform3D()) = 0;
  virtual void Draw2D (const G4Circle&,
    const G4Transform3D& objectTransformation = G4Transform3D()) = 0;
  virtual void Draw2D (const G4Polyhedron&,
    const G4Transform3D& objectTransformation = G4Transform3D()) = 0;
  virtual void Draw2D (const G4Polyline&,
    const G4Transform3D& objectTransformation = G4Transform3D()) = 0;
  virtual void Draw2D (const G4Polymarker&,
    const G4Transform3D& objectTransformation = G4Transform3D()) = 0;
  virtual void Draw2D (const G4Square&,
    const G4Transform3D& objectTransformation = G4Transform3D()) = 0;
  virtual void Draw2D (const G4Text&,
    const G4Transform3D& objectTransformation = G4Transform3D()) = 0;
  virtual void Draw (const G4VHit&) = 0;
  virtual void Draw (const G4VDigi&) = 0;
  virtual void Draw (const G4VTrajectory&) = 0;
  virtual void Draw (const G4LogicalVolume&, const G4VisAttributes&,
    const G4Transform3D& objectTransformation = G4Transform3D()) = 0;
  virtual void Draw (const G4VPhysicalVolume&, const G4VisAttributes&,
    const G4Transform3D& objectTransformation = G4Transform3D()) = 0;
  virtual void Draw (const G4VSolid&, const G4VisAttributes&,
    const G4Transform3D& objectTransformation = G4Transform3D()) = 0;
};
\end{boxedverbatim}
}
\end{center}
\caption{The basic {\tt G4VVisManager} interface.}
\label{figG4VVisManager}
\end{figure}

The Geant4 visualization manager interface is available to all Geant4 code whether in a \Gfour user application or in the toolkit itself.  However, the coder must check that a concrete implementation actually exists and avoid using the interface if not; for example:

{\footnotesize
\begin{verbatim}
  G4VVisManager* pVVisManager = G4VVisManager::GetConcreteInstance();
  if (pVVisManager) {
    pVVisManager->Draw(circle);
    ...
\end{verbatim}
}

This allows one to build a \Gfour application without a concrete implementation, for example for batch production, even if such code remains in the application or the toolkit (which it certainly does).

A second interface---{\tt G4VGraphicsScene} (Figure \ref{figG4VGraphicsScene})---is private to the toolkit.  It is not available for application developers. It is used by the toolkit---geometry and modeling in particular. The \vm is expected to initiate its use by providing a reference to a concrete implementation.

\begin{figure}
\begin{center}
{\scriptsize
\begin{boxedverbatim}
class G4VGraphicsScene {
public:
  G4VGraphicsScene();
  virtual ~G4VGraphicsScene();
  virtual void PreAddSolid (const G4Transform3D& objectTransformation,
                            const G4VisAttributes& visAttribs) = 0;
  virtual void PostAddSolid () = 0;
  virtual void AddSolid (const G4Box&)       = 0;
  virtual void AddSolid (const G4Cons&)      = 0;
  virtual void AddSolid (const G4Tubs&)      = 0;
  virtual void AddSolid (const G4Trd&)       = 0;
  virtual void AddSolid (const G4Trap&)      = 0;
  virtual void AddSolid (const G4Sphere&)    = 0;
  virtual void AddSolid (const G4Para&)      = 0;
  virtual void AddSolid (const G4Torus&)     = 0;
  virtual void AddSolid (const G4Polycone&)  = 0;
  virtual void AddSolid (const G4Polyhedra&) = 0;
  virtual void AddSolid (const G4VSolid&)    = 0;  // For solids not above.
  virtual void AddCompound (const G4VTrajectory&)        = 0;
  virtual void AddCompound (const G4VHit&)               = 0;
  virtual void AddCompound (const G4VDigi&)              = 0;
  virtual void AddCompound (const G4THitsMap<G4double>&) = 0;
  virtual void BeginPrimitives
  (const G4Transform3D& objectTransformation = G4Transform3D()) = 0;
  virtual void EndPrimitives () = 0;
  virtual void BeginPrimitives2D
  (const G4Transform3D& objectTransformation = G4Transform3D()) = 0;
  virtual void EndPrimitives2D () = 0;
  virtual void AddPrimitive (const G4Polyline&)   = 0;
  virtual void AddPrimitive (const G4Scale&)      = 0;
  virtual void AddPrimitive (const G4Text&)       = 0;
  virtual void AddPrimitive (const G4Circle&)     = 0;
  virtual void AddPrimitive (const G4Square&)     = 0;
  virtual void AddPrimitive (const G4Polymarker&) = 0;
  virtual void AddPrimitive (const G4Polyhedron&) = 0;
};
\end{boxedverbatim}
}
\end{center}
\caption{The {\tt G4VGraphicsScene} low-level interface.}
\label{figG4VGraphicsScene}
\end{figure}

In summary, a \vs is required to implement \emph{two} interfaces---{\tt G4VVisManager} (Figure \ref{figG4VVisManager}) and {\tt G4VGraphicsScene} (Figure \ref{figG4VGraphicsScene}).  The latter is, in fact, intended to represent the ``scene handler''. Objects that are passed to it ({\tt G4Box} objects, etc.) are required to be turned into visible renderings.  Quite how this is done is up to the \vs.

\section{The \GVS}

So now we are in a a position to describe the \GVS, which is the implementation of the above interfaces that is distributed with the \Gfour toolkit.

\begin{figure}
\centerline{\includegraphics[width=\linewidth]{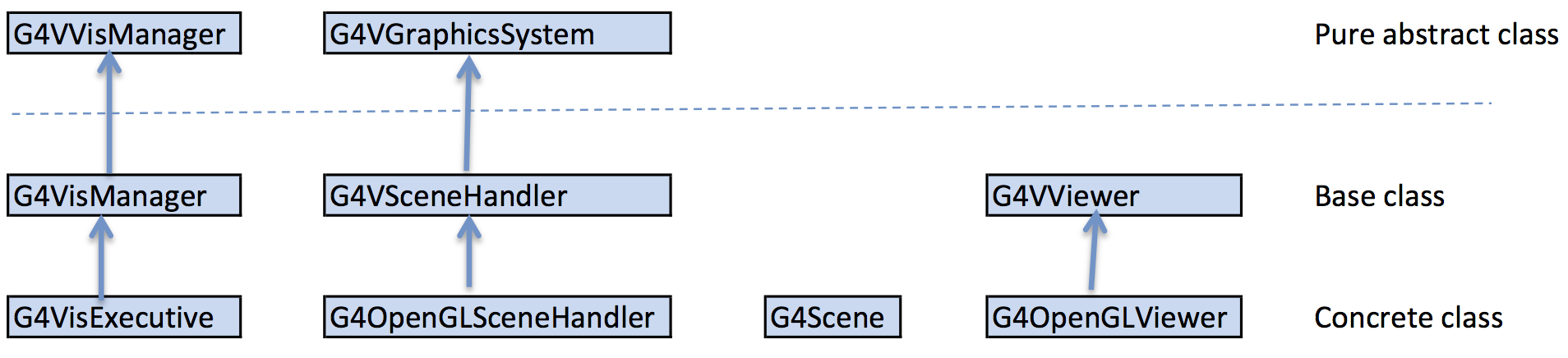}}
\vspace*{8pt}
\caption{The basic class diagram.  OpenGL is shown as an example; in fact the system is capable of handling multiple scene handlers and viewers.}
\label{figBasicClassDiagram}
\end{figure}

\Gfour defers to the application developer the decision of which external graphics packages should be required. Accordingly, it is the application developer's responsibility to declare which graphics drivers should be made available to the end user. He or she makes these declarations by instantiating a {\tt G4VisManager} (written to the {\tt G4VVisManager} interface) and registering appropriate graphics drivers. The \Gfour distribution provides a general purpose sub-class called {\tt G4VisExecutive}, which collaborates with the \Gfour build system to offer all available drivers and which is used in all \Gfour examples. This is shown in Figure \ref{figBasicClassDiagram}.

{\tt G4VSceneHandler} is written to the {\tt G4VGraphicsSystem} interface and itself is the base class for graphics-library dependent concrete scene handlers, for example, for OpenGL.  The system is capable of handling multiple drivers, i.e., multiple scene handlers and viewers.

\subsection{Visualisation of touchables}

\Gfour has a layered geometry structure, in which ÒsolidsÓ define shapes, Òlogical volumesÓ add material information to solids and Òphysical volumesÓ place a given logical volume in space. To make efficient use of memory, Geant4 provides mechanisms whereby a single physical volume may have more than one placement, using mechanisms called Replica placement or Parameterised placement. Whether a physical volume has only one placement, or many placements, each placement is called a touchable.

The \vs must deal in touchables, and it is the job of {\tt G4PhysicalVolumeModel} to roll out the \Gfour geometry into touchables.  To avoid being overwhelmed, it is possible to specify starting points in the geometry hierarchy, such as specific sub-detectors, and limit the depth of descent so as to avoid too much detail.  One can also control, for example, the visibility with {\tt G4VisAttributes} (Section \ref{secAttributes}); invisible touchables may then be suppressed (culled).

It is possible to edit the {\tt G4VisAttributes} of logical volumes with {\tt /vis/geometry/} commands and to modify the {\tt G4VisAttributes} of individual touchables with {\tt /vis/touchable/} commands.  The latter can also be done interactively in the OpenGL Qt viewer.

\subsection{Visualisation of transients}
\label{secTransients}

``Transients'' is a term for entities that appear for a limited time on a ``permanent'' background, for example, particle trajectories in a detector. The \GVS distinguishes these; smart drivers may take advantage. For example the OpenGL ``stored'' system (Section \ref{secOpenGL}) keeps separate lists in its graphical database for transient and permanent objects, and the \GVM clears and rebuilds the former without having to rebuild the latter. This makes is very efficient for displaying event after event on a fixed detector image.

The \GVS allows one to display particle trajectories, hits and digits event by event or to accumulate them, one on the other, until the end of run.

This is fine for drivers with their own graphical database, but in such circumstances, not-so-smart drivers have to re-draw both permanent and transient objects, with a noticeable performance degradation. Moreover they may not even have the ability to remember permanent objects, let alone transients. For these drivers the \GVS keeps a memory of permanent objects (for example, it may always re-visit the geometry hierarchy to rebuild a detector image) and also keeps a memory, to a limited but considerable extent, of transient objects such as particle trajectories, hits and digits that are stored in simulated events. With this feature, described in Section \ref{secEventStoring}, all drivers, including the not-so-smart, can recover the transients and emulate the superposition of transients on permanents, a switch of viewpoint or even a switch of drivers.

Transients that are generated from user code will not be recoverable in this way, unless the user encloses the code in a User Vis Action--see Section \ref{secUserVisAction}.

\subsection{Event storing}
\label{secEventStoring}

The \GVM asks the Run Manager to keep events so that they may be accessed to redraw, say, the particle trajectories, hits and digits in a view.  This is particularly useful for drivers that do not have their own graphical database or when switching from one driver to another.  The default behaviour is to keep the last 100 events, but the number may be adjusted to take into account available memory; also user code can provide algorithms to specify which events should be kept.

\subsection{Trajectory modelling and filtering}

An extensive set of models and filters for trajectories drawing is available.  They control the colour and visibility by particle type and the visibility by momentum or detector volume, for example.  Some commands are part of the start-up script shown in Figure \ref{figVisMac}. Models may also select whether the trajectory is drawn as a line, as step points, or as both.

A similar but less extensive set is available also for user-defined hits and digits.  A user may add attributes to trajectories, hits and digits and filter them at drawing time with {\tt /vis/filtering/} commands.

\subsection{Attributes}
\label{secAttributes}

Every drawable entity is assigned visualisation attributes {\tt G4VisAttributes} either by the user or from a modifiable default set.  They include visibility (i.e., drawn or not if culling is active), colour, line width, line style, and the ability to ``force'' some modes of drawing, such as wireframe or with surfaces, regardless of the general user request.

Some entities have additional attributes in the form of {\tt G4AttDef} and {\tt G4AttValue} pairs, which are strings to be interpreted as text, integers, doubles or three-vectors, with or without dimension (units), following the HepRep attribute design.\cite{HepRep}  For example, trajectories acquire particle type, momentum, volume name, process type.  A full list for {\tt G4RichTrajectory} objects is shown in Figure \ref{figRichTrajectory}.  (Note that this is memory-consuming; by default trajectories as stored as {\tt G4Trajectory} objects with fewer attributes.)

\begin{figure}
\begin{center}
{\tiny
\begin{boxedverbatim}
/G4TrajectoriesModel:
  Event ID (EventID): G4int
G4RichTrajectory:
  Creator Process Name (CPN): G4String
  Creator Process Type Name (CPTN): G4String
  Charge (Ch): unit: e+ (G4double)
  Ending Process Name (EPN): G4String
  Ending Process Type Name (EPTN): G4String
  Final kinetic energy (FKE): G4BestUnit (G4double)
  Final Next Volume Path (FNVPath): G4String
  Final Volume Path (FVPath): G4String
  Track ID (ID): G4int
  Initial kinetic energy (IKE): G4BestUnit (G4double)
  Initial momentum magnitude (IMag): G4BestUnit (G4double)
  Initial momentum (IMom): G4BestUnit (G4ThreeVector)
  Initial Next Volume Path (INVPath): G4String
  Initial Volume Path (IVPath): G4String
  No. of points (NTP): G4int
  PDG Encoding (PDG): G4int
  Parent ID (PID): G4int
  Particle Name (PN): G4String
G4RichTrajectoryPoint:
  Auxiliary Point Position (Aux): G4BestUnit (G4ThreeVector)
  Process Defined Step (PDS): G4String
  Process Type Defined Step (PTDS): G4String
  Position (Pos): G4BestUnit (G4ThreeVector)
  Post-step-point status (PostStatus): G4String
  Post-step-point global time (PostT): G4BestUnit (G4double)
  Post-step Volume Path (PostVPath): G4String
  Post-step-point weight (PostW): G4double
  Pre-step-point status (PreStatus): G4String
  Pre-step-point global time (PreT): G4BestUnit (G4double)
  Pre-step Volume Path (PreVPath): G4String
  Pre-step-point weight (PreW): G4double
  Remaining Energy (RE): G4BestUnit (G4double)
  Total Energy Deposit (TED): G4BestUnit (G4double)
\end{boxedverbatim}
}
\end{center}
\caption{The attributes of a {\tt G4RichTrajectory} object.}
\label{figRichTrajectory}
\end{figure}

Touchables also get attributes---see Figure \ref{figTouchableAttributes}.

\begin{figure}
\begin{center}
{\tiny
\begin{boxedverbatim}
G4PhysicalVolumeModel:
  Material Density (Density): G4BestUnit (G4double)
  Dump of Solid properties (DmpSol): G4String
  Entity Type (EType): G4String
  Logical Volume (LVol): G4String
  Material Name (Material): G4String
  Physical Volume Path (PVPath): G4String
  Material Radiation Length (Radlen): G4BestUnit (G4double)
  Cuts Region (Region): G4String
  Root Region (0/1 = false/true) (RootRegion): G4bool
  Solid Name (Solid): G4String
  Material State (enum undefined,solid,liquid,gas) (State): G4String
  Transformation of volume (Trans): G4String
\end{boxedverbatim}
}
\end{center}
\caption{The attributes of touchables provided by {\tt G4PhysicalVolumeModel}.}
\label{figTouchableAttributes}
\end{figure}

It is these attributes that are selectable with {\tt /vis/filtering/} commands, as mentioned above, and that may be dumped by picking with pick-sensitive drivers---OpenGL, OpenInventor and HepRApp (the browser for HepRepFile)---see below.

\subsection{User Vis Actions}
\label{secUserVisAction}

As we have said, a user may draw to the current viewer at any time with C++ code written to the basic abstract interface (Section \ref{secAbstractInterface}).  However, the drawing will belong to the set of unrecoverable ``transients'' (see Section \ref{secTransients}) and will not be preserved on change of view or change of driver.  A better strategy is to write a User Vis Action, a class that inherits {\tt G4UserVisAction}, instantiate it and register it with the \GVM.  This way the \vm can invoke the code repeatedly as required to give the drawn objects as sort of permanence.

A few examples are included in the \Gfour distribution.  {\tt examples/extended/visualization/userVisAction} shows how to implement a user-defined logo.  {\tt examples/extended/visualization/standalone} shows, as the name implies, how the \Gfour libraries can be used as a ``standalone'' multi-driver graphics library.  Figure \ref{figUserVisAction} shows the code that results in Figure \ref{figStandalone}.  The reader will notice that one may use the \Gfour geometry, including, for example, the Boolean solids ({\tt G4SubtractionSolid}, etc.).

\begin{figure}
\begin{center}
{\scriptsize
\begin{boxedverbatim}
class StandaloneVisAction: public G4VUserVisAction {
  virtual void Draw();
};
void StandaloneVisAction::Draw() {
  G4VVisManager* pVisManager = G4VVisManager::GetConcreteInstance();
  if (pVisManager) {
    // Simple box...
    pVisManager->Draw(G4Box("box",2*m,2*m,2*m),
                      G4VisAttributes(G4Colour(1,1,0)));
    // Boolean solid...
    G4Box boxA("boxA",3*m,3*m,3*m);
    G4Box boxB("boxB",1*m,1*m,1*m);
    G4SubtractionSolid subtracted("subtracted_boxes",&boxA,&boxB,
                       G4Translate3D(3*m,3*m,3*m));
    pVisManager->Draw(subtracted,
                      G4VisAttributes(G4Colour(0,1,1)),
                      G4Translate3D(-6*m,-6*m,-6*m));
  }
}
\end{boxedverbatim}
}
\end{center}
\caption{A User Vis Action that gives drawn objects a permanence.}
\label{figUserVisAction}
\end{figure}

\begin{figure}
\centerline{\includegraphics[width=\linewidth]{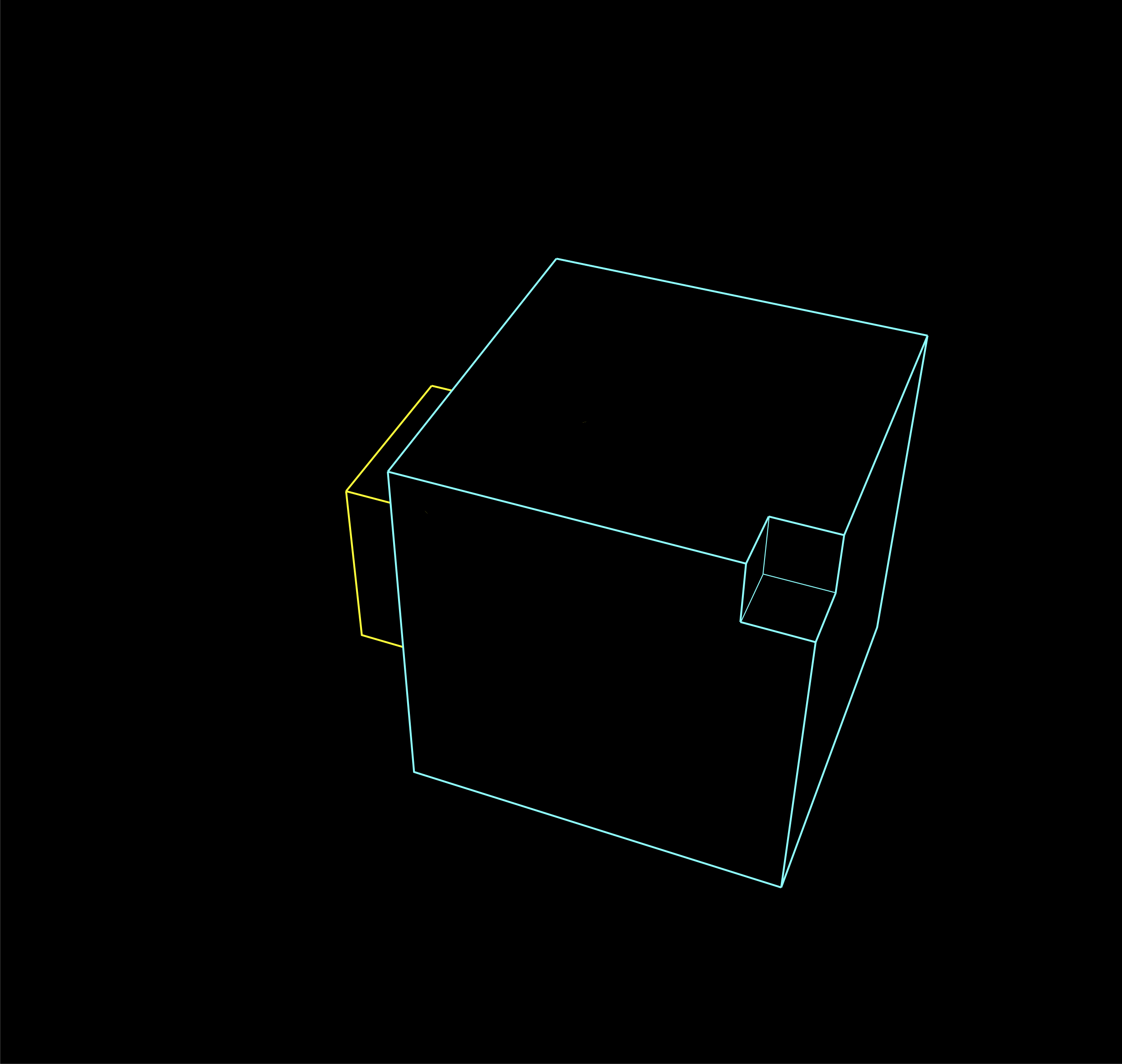}}
\vspace*{8pt}
\caption{A result of the code shown in Figure \ref{figUserVisAction}.  (This figure was produced with {\tt /vis/ogl/printEPS} from an OpenGL stored Xm window.)}
\label{figStandalone}
\end{figure}

\section{The User Interface}
\label{secUI}

A user may code to the basic abstract interface, {\tt G4VVisManager}, or, more usually, use an extensive set of visualisation commands (Section \ref{secVisCommands}) through the \Gfour User Interface.  \Gfour offers several user interfaces, ranging from dumb terminal to graphical.  In general, any user interface will work with any visualisation driver, except in the case of Qt\cite{Qt} (which offers the most advanced interactivity, and for which the graphics window is closely coupled to the user interface).

A user may also read commands from a file by typing ``{\tt /control/execute <filename>}'' or by programming the reading of a file using {\tt G4UImanager::ApplyCommand()}. 
This is typically to be found in the examples distributed with the \Gfour toolkit.

\subsection{Visualisation commands}
\label{secVisCommands}

The visualisation commands are too numerous to detail here.  A full description of all commands is available in the Application Developers Guide.\cite{UGA}

For example, a user would instantiate a user interface and then issue commands, such as ``{\tt /vis/open OGL}'' and ``{\tt /vis/drawVolume}'' to get an image of the detector.  Figure \ref{figVisMac} shows a typical start-up script.  After reading this file, ``{\tt /run/beamOn 10}'', for example, would run the simulation and draw particle trajectories from 10 events.

\begin{figure}
\begin{center}
{\tiny
\begin{boxedverbatim}
/vis/open OGL 600x600-0+0

# Disable auto refresh and quieten vis messages whilst scene and
# trajectories are established:
/vis/viewer/set/autoRefresh false
/vis/verbose errors

/vis/drawVolume

# Specify view angle:
/vis/viewer/set/viewpointVector -1 0 0
/vis/viewer/set/lightsVector -1 0 0

# Specify style (surface, wireframe, auxiliary edges,...)
/vis/viewer/set/style wireframe
/vis/viewer/set/auxiliaryEdge true
/vis/viewer/set/lineSegmentsPerCircle 100

# Draw smooth trajectories at end of event, showing trajectory points
# as markers 2 pixels wide:
/vis/scene/add/trajectories smooth
/vis/modeling/trajectories/create/drawByCharge
/vis/modeling/trajectories/drawByCharge-0/default/setDrawStepPts true
/vis/modeling/trajectories/drawByCharge-0/default/setStepPtsSize 2

# To draw only gammas:
#/vis/filtering/trajectories/create/particleFilter
#/vis/filtering/trajectories/particleFilter-0/add gamma

# To superimpose all of the events from a given run:
/vis/scene/endOfEventAction accumulate

# DecorationsÉ
# Name
/vis/set/textColour green
/vis/scene/add/text2D 0 -.9 24 ! ! exampleB1
/vis/set/textLayout    # Revert to normal (left adjusted) layout
/vis/set/textColour    # Revert to default text colour (blue)

# Axes, scale, etc.
/vis/scene/add/scale   # Simple scale line
/vis/scene/add/axes    # Simple axes: x=red, y=green, z=blue.
/vis/scene/add/eventID # Drawn at end of event
/vis/scene/add/date    # Date stamp
/vis/scene/add/logo2D  # Simple logo
/vis/scene/add/logo    # 3D logo

# Frame
/vis/set/colour red
/vis/set/lineWidth 2
/vis/scene/add/frame   # Simple frame around the view
/vis/set/colour        # Revert to default colour (white)
/vis/set/lineWidth     # Revert to default line width (1.)

# Attach text to one edge of Shape1, with a small, fixed offset
/vis/scene/add/text 0 6 -4 cm 18 4 4 Shape1
# Attach text to one corner of Shape2, with a small, fixed offset
/vis/scene/add/text 6 7 10 cm 18 4 4 Shape2

# To get nice view
/vis/geometry/set/visibility World 0 false
/vis/geometry/set/visibility Envelope 0 false
/vis/viewer/set/style surface
/vis/viewer/set/hiddenMarker true
/vis/viewer/set/viewpointThetaPhi 120 150

# Re-establish auto refreshing and verbosity:
/vis/viewer/set/autoRefresh true
/vis/verbose warnings

# For file-based drivers, use this to create an empty detector view:
#/vis/viewer/flush
\end{boxedverbatim}
}
\end{center}
\caption{A typical start-up script}
\label{figVisMac}
\end{figure}

\section{The Drivers}

Over the years we have developed, at the latest count, 14 drivers of various sorts (or upwards of 20 if one counts all the OpenGL variants and DAWN and VRML variants).  A user may draw to the basic abstract interfaces, either in C++ code or, more usually, via
visualisation commands (Section \ref{secVisCommands}) through a user interface, and expect it to be rendered in one of a number of different ways: to a computer screen (graphics drivers, Section \ref{secGraphicsDrivers}); or to a file for subsequent browsing (file-writing drivers, Section \ref{secFileWriters}.  There is also a category of pseudo-drivers (Section \ref{secPseudo}).

Some drivers support picking, i.e., clicking on an item in the graphics window pops up a window of information or dumps a print-out of attributes to standard output.  In this category are: HepRepFile; OpenGL X11; OpenGL Qt; Open Inventor.

\subsection{Graphics drivers}
\label{secGraphicsDrivers}

The workhorse of the \GVS is the set of OpenGL drivers.  We also have a graphics driver for Open Inventor and a driver, RayTracer, that uses Geant4's own tracking algorithms to produce a ray-traced image.

\subsubsection{OpenGL}
\label{secOpenGL}

A particular feature of OpenGL is that one may store GL commands in a ``display list'' that may be efficiently rendered by a graphics processing unit.  We recommend this ``stored mode'' as the default option to get good performance for rotating, zooming, etc., without having to re-visit the \Gfour kernel.  However a complex detector or complex event structure can overwhelm a computer system, so we offer an ``immediate mode'' whereby objects are rendered directly to the screen, but to change viewpoint, for example, the graphics system re-visits the \Gfour kernel for information about the scene and makes all the coordinate calculations afresh, which is much slower.

\begin{figure}
\centerline{\includegraphics[width=\linewidth]{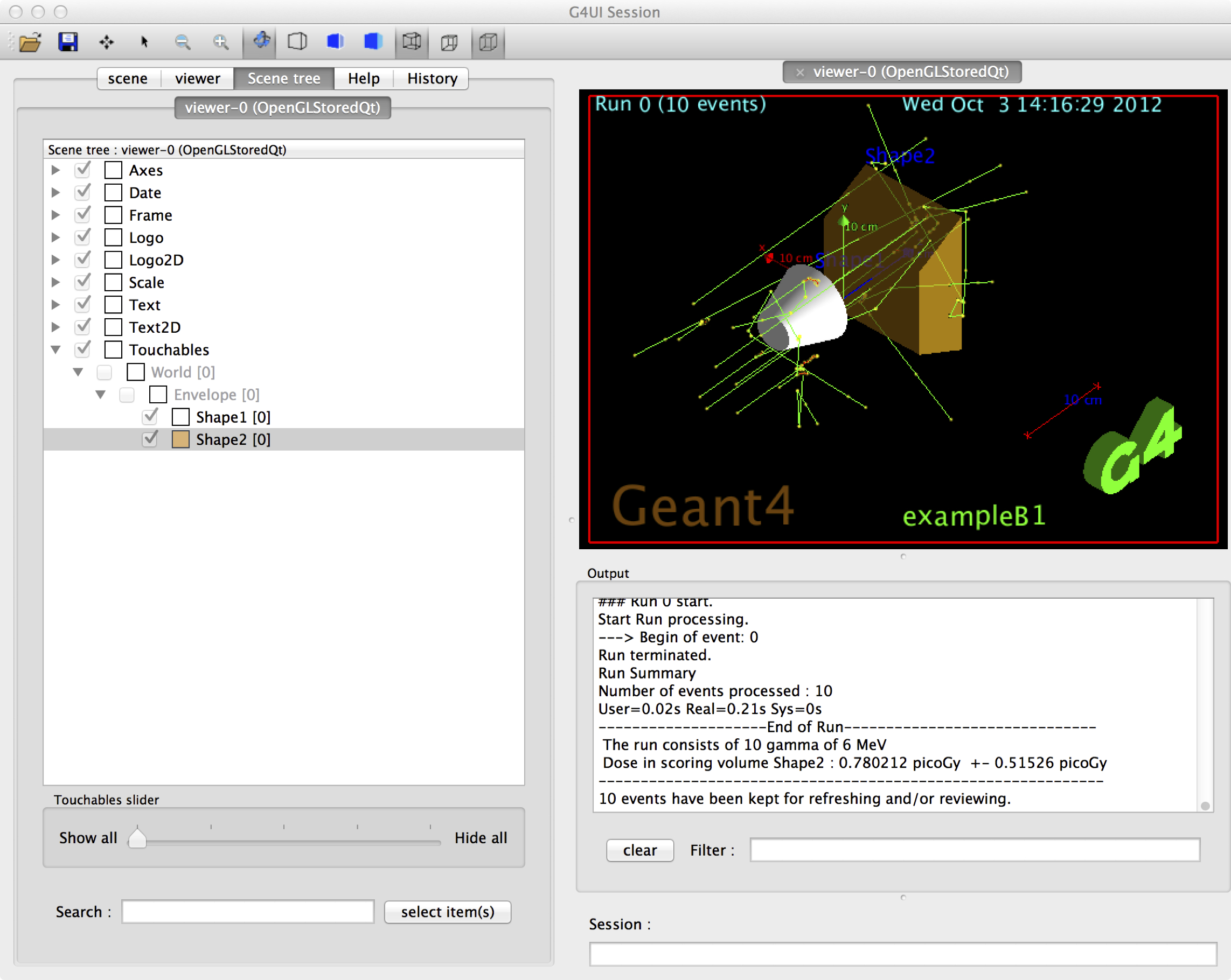}}
\vspace*{8pt}
\caption{A screen shot of an interactive Qt session.}
\label{figQt}
\end{figure}

The OpenGL drivers have various degrees of interactivity depending on the availability of advanced graphics libraries.  The simplest---OpenGL X11 (Unix) and Win32 (Microsoft Windows)--- are passive.  For example, to change the viewpoint one must issue a command, e.g., ``{\tt /vis/viewer/set/viewpointThetaPhi 30 30 deg}''.  With Motif\cite{bibOpenMotif} libraries one can build OpenGL Xm; the viewer provides some interactivity via pull-down menus, including rotation and zoom.  The most sophisticated is the OpenGL Qt\cite{Qt} driver, which offers a huge amount of interactivity, including rotation, pan and zoom, picking, drawing style, projection style, etc.  The Qt user interface, {\tt G4UIQt}, which must be used with it, includes an interactive help system and an interactive portrayal of the scene, including the geometry hierarchy, through which one can change the colour and visibility of individual screen objects.  Figure \ref{figQt} shows a typical screen shot.

The OpenGL driver also implements transparency and cutaways.

A user may save the view to file in EPS format with {\tt /vis/ogl/printEPS}.

\subsubsection{Open Inventor}

\begin{figure}
\centerline{\includegraphics[width=\linewidth]{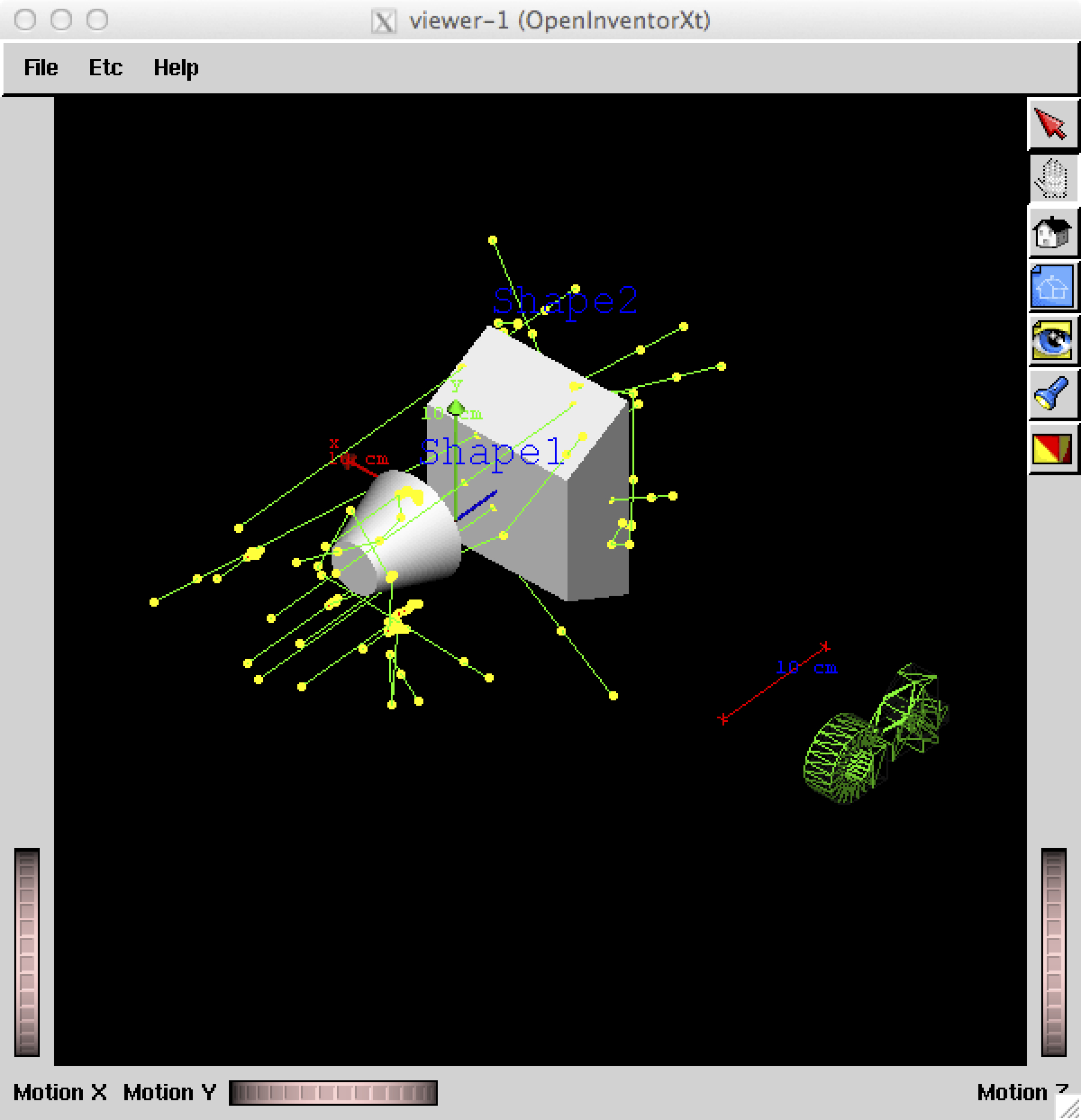}}
\vspace*{8pt}
\caption{A screen shot of an Open Inventor viewer.}
\label{figOI}
\end{figure}

The Open Inventor drivers for Xt (Unix) and Win32 (Microsoft Windows) also provides good interactivity.  Figure \ref{figOI} shows a typical screen shot.  At present this driver cannot  render ``2D'' objects such as non-moving descriptive text.

It may be worth mentioning that as in OpenGL the view may be saved in EPS format via a menu button on the viewer

\subsubsection{Ray Tracer for X}
\label{secRayTarcerX}
\begin{figure}
\centerline{\includegraphics[width=\linewidth]{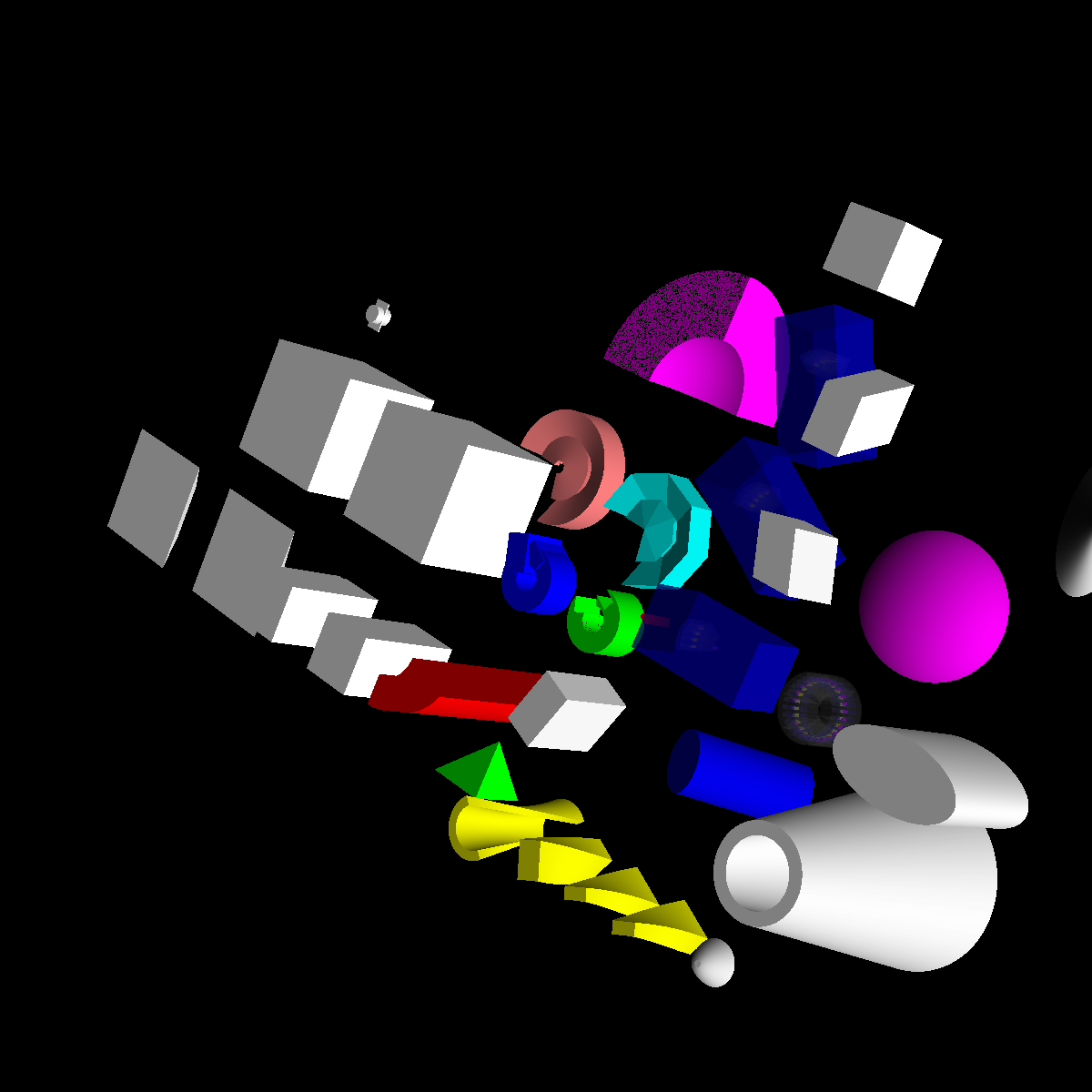}}
\vspace*{8pt}
\caption{A screen shot of a Ray Tracer viewer for X11.}
\label{figRayTracerX}
\end{figure}

The Ray Tracer can only render geometry.  Figure \ref{figRayTracerX} shows a typical screen shot.  The Ray Tracer can also produce JPEG files directly---see Section \ref{secRTjpeg}.

\subsection{File-writing drivers}
\label{secFileWriters}

\subsubsection{HepRepFile}

HepRepFile allows the current scene (geometry, event or collections of events) to be rendered to an XML file in the HepRep format.\cite{HepRep} The file can then be read into a HepRep browser such as HepRApp.\cite{HepRApp}
Because the HepRep file contains a full 3D description of the scene, augmented with full physics attribute data (attributes of the touchables, trajectories and hits), the HepRApp user can then rotate, pan, zoom, change projection styles, pick to view attributes and filter graphics based on attributes. Resulting images can be exported a wide variety of graphics formats (gif, pdf, etc.).

\begin{figure}
\centerline{\includegraphics[width=\linewidth]{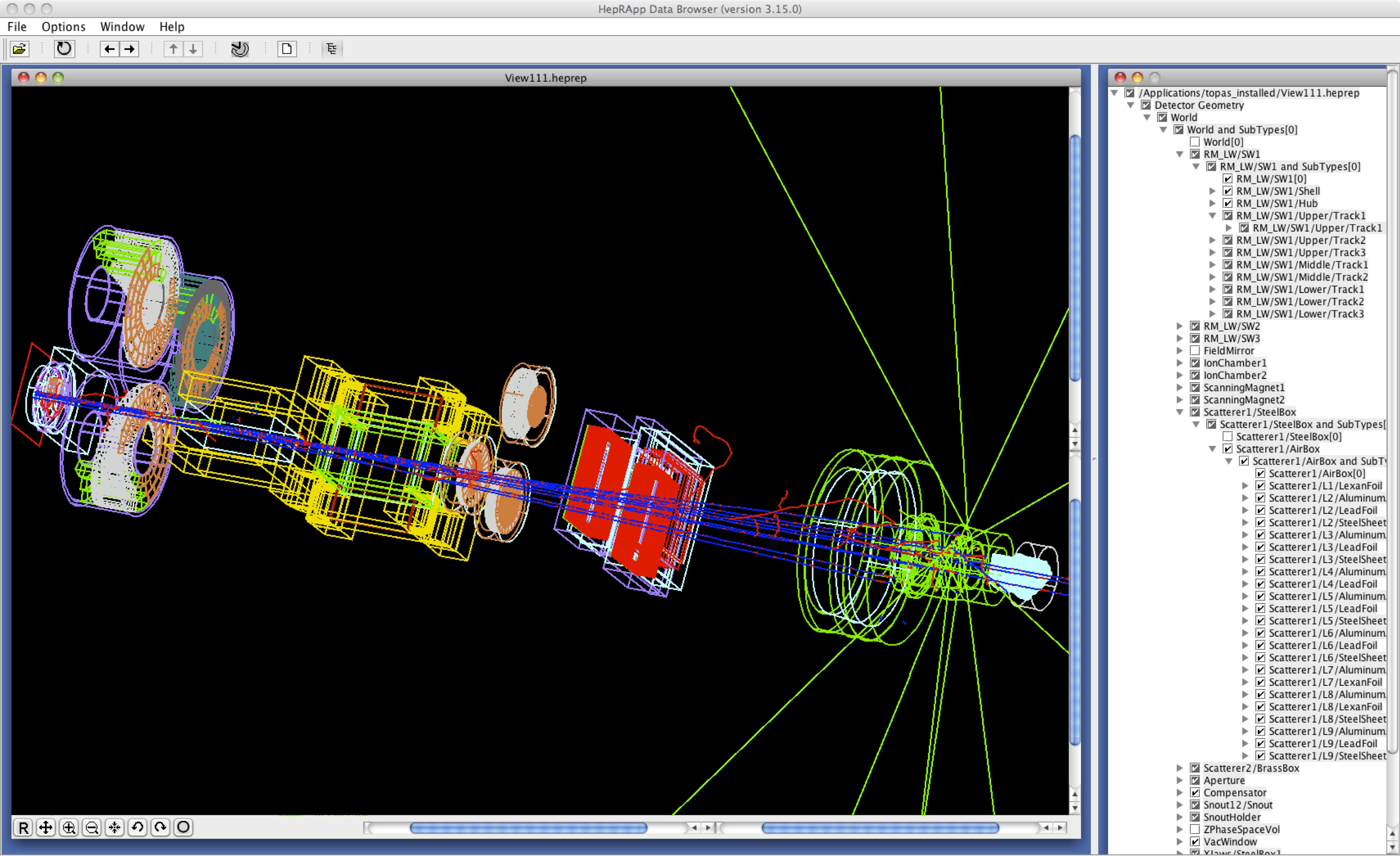}}
\vspace*{8pt}
\caption{A screen shot of HepRApp browsing a file produced by HepRepFile.}
\label{figHepRepFile}
\end{figure}

 Figure \ref{figHepRepFile} shows a screen shot of HepRApp browsing a file produced by HepRepFile.

\subsubsection{DAWNFILE}

\begin{figure}
\centerline{\fbox{\includegraphics[width=\linewidth]{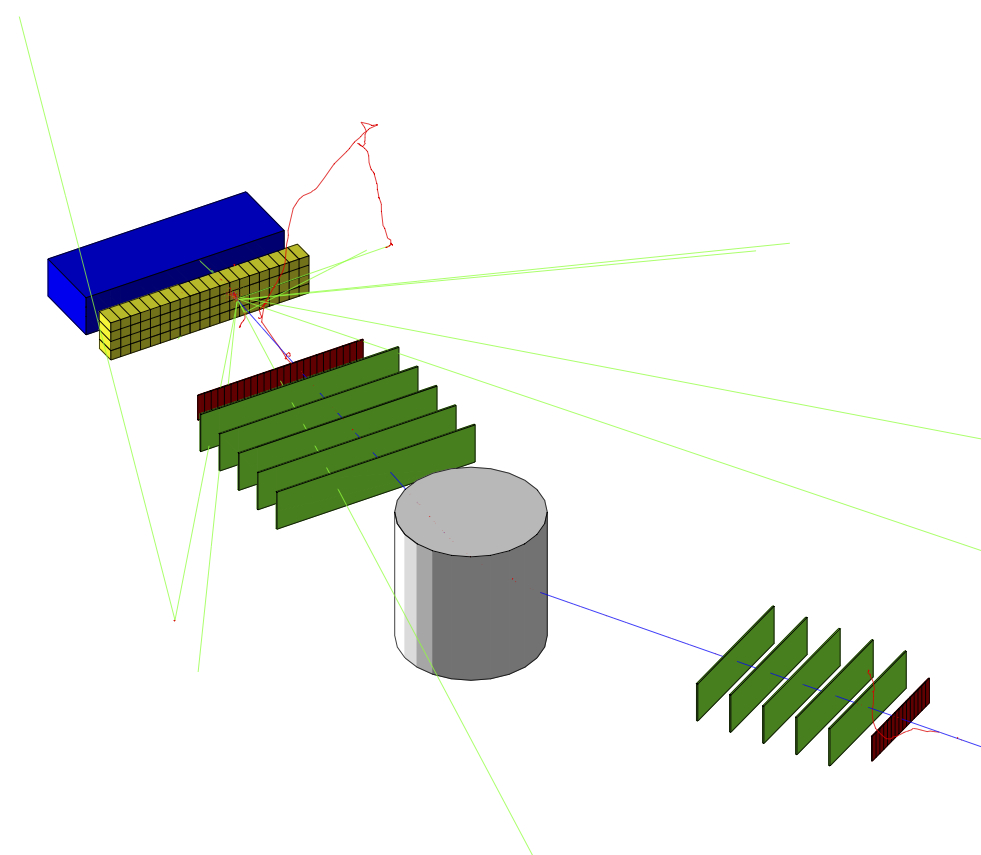}}}
\vspace*{8pt}
\caption{A rendering of DAWNFILE output with DAWN.}
\label{figDAWN}
\end{figure}

Similarly, DAWNFILE produces a file suitable for browsing with DAWN.\cite{VisCHEP97DAWN} DAWN is capable of sophisticated hidden line and hidden surface removal and produces very high quality images in encapsulated PostScript format--see Figure \ref{figDAWN}.

One can generate cutaways with DAWNCUT.\cite{VisDAWNCUT} A useful spin-off is DAVID,\cite{VisDAVID} which is a volume overlap detection application.

\subsubsection{VRMLFILE}

Similarly, VRML2FILE produces a file suitable for browsing with a VRML browser.  Note that VRML1 is still available.

\begin{figure}
\centerline{\includegraphics[width=\linewidth]{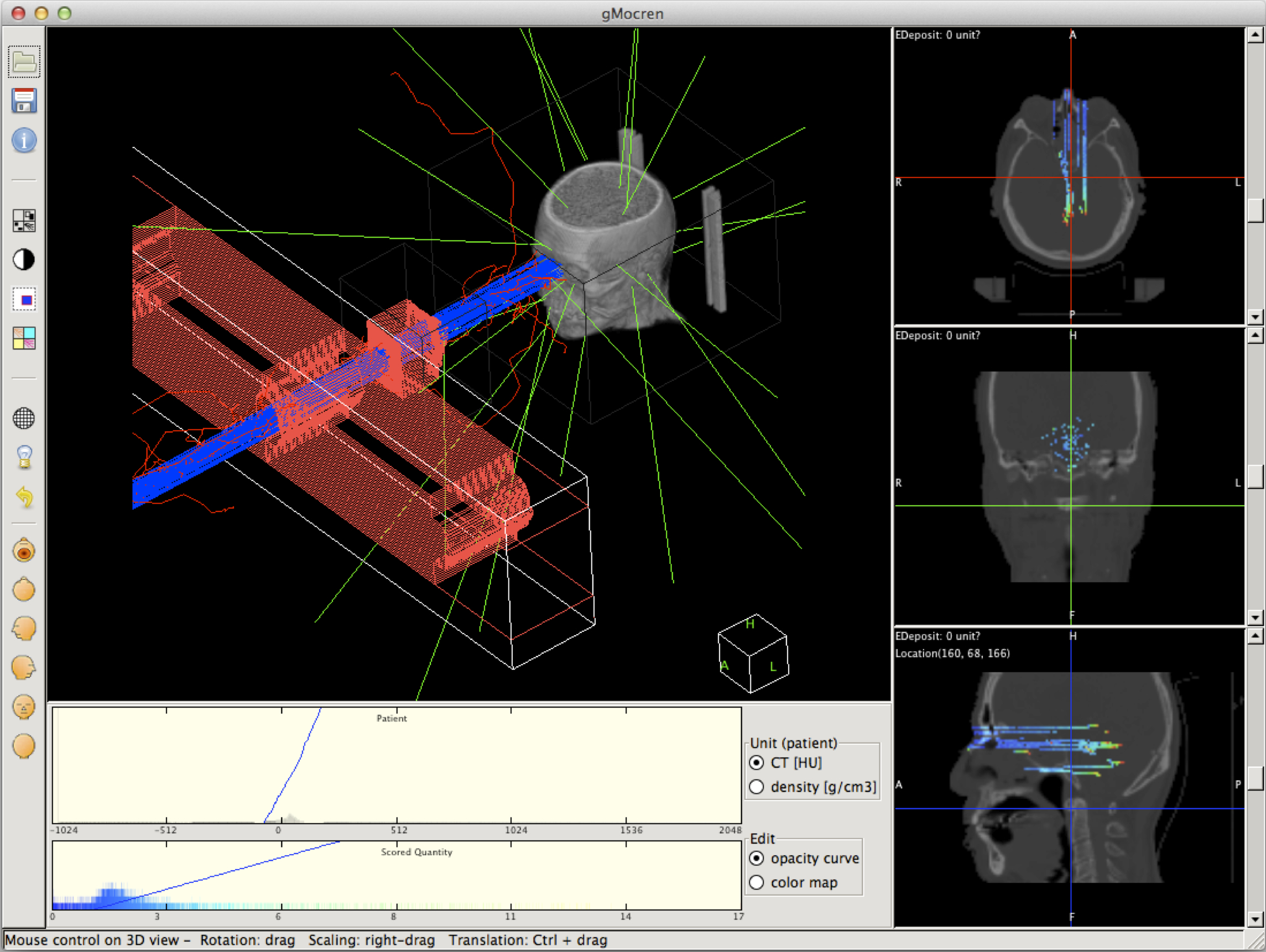}}
\vspace*{8pt}
\caption{A gMocren rendering of a dose distribution produced with gMocrenFile.}
\label{figgMocren}
\end{figure}

\subsubsection{gMocrenFile}

gMocren\cite{bibgMocren} is used typically to visualise radiation therapy dose data.  Figure \ref{figgMocren} shows a rendering with a gMocren viewer of a file that has been created with gMocrenFile after a simulation run. 

\subsubsection{Ray Tracer jpeg-writer}
\label{secRTjpeg}

The Ray Tracer can produce a JPEG file, either from its X11 version (described above, Section \ref{secRayTarcerX}, and instantiated with ``{\tt /vis/open RayTracerX}'') or from this version that needs no graphics library, instatiated within ``{\tt /vis/open RayTracer}''.

\subsubsection{Encapsulated PostScript}

For completeness we re-iterate the possibility, mentioned above, of saving a view with any variant of OpenGL or Open Inventor driver to file in EPS format with the visualisation command {\tt /vis/ogl/printEPS}.  This can be done even without a user interface (see Section \ref{secUI}) by using the {\tt ApplyCommand()} method of the User Interface Manager. The DAWNFILE and HepRepFile drivers can also be used in this way to produce files that can be browsed to produce an EPS file.

\subsection{BSD Socket drivers}

There are versions of the DAWN and VRML drivers that can communicate with their respective browsers though a socket mechanism.  The browsers have to be launched first and suitable socket numbers chosen.

\subsection{Pseudo-drivers}
\label{secPseudo}

We have taken advantage of the way the \vs works to write a useful pseudo-driver.  The \vs renders the scene to the ``Tree'' driver and dumps useful information.

\subsubsection{ASCII Tree}

At the moment this is the only tree driver.  It dumps the geometry tree to standard output and lists the physical volume name and optionally also the logical volume name, the solid name, its volume, its density, its mass excluding daughters and the total mass of the top physical volume.  Figure \ref{figASCIITree} shows typical output.

\begin{figure}
\begin{center}
{\scriptsize
\begin{boxedverbatim}
G4ASCIITreeSceneHandler::BeginModeling: writing to G4 standard output (G4cout)
#  Set verbosity with "/vis/ASCIITree/verbose <verbosity>":
#    <  10: - does not print daughters of repeated placements, does not repeat replicas.
#    >= 10: prints all physical volumes.
#  The level of detail is given by verbosity%10:
#  for each volume:
#    >=  0: physical volume name.
#    >=  1: logical volume name (and names of sensitive detector and readout geometry, if any).
#    >=  2: solid name and type.
#    >=  3: volume and density.
#    >=  5: daughter-subtracted volume and mass.
#  and in the summary at the end of printing:
#    >=  4: daughter-included mass of top physical volume(s) in scene to depth specified.
#  Note: by default, culling is switched off so all volumes are seen.
#  Note: the mass calculation takes into account daughters, which can be time consuming.  If
#    you want the mass of a particular subtree to a particular depth:
#    /vis/open ATree
#    /vis/ASCIITree/verbose 14
#    /vis/scene/create
#    /vis/scene/add/volume <subtree-physical-volume> ! <depth>
#    /vis/sceneHandler/attach
#    /vis/viewer/flush
#  Now printing with verbosity 15
#  Format is: PV:n / LV (SD,RO) / Solid(type), volume, density, daughter-subtracted volume and mass
#  Abbreviations: PV = Physical Volume,     LV = Logical Volume,
#                 SD = Sensitive Detector,  RO = Read Out Geometry.
"World":0 / "World" / "World"(G4Box), 20736 cm3, 1.20479 mg/cm3 (G4_AIR), 8736 cm3, 10.525 g 
    "Envelope":0 / "Envelope" / "Envelope"(G4Box), 12000 cm3, 1 g/cm3  (G4_WATER), 10888.1 cm3, 10.8881 kg
      "Shape1":0 / "Shape1" / "Shape1"(G4Cons), 175.929 cm3, 1.127 g/cm3  (G4_A-150_TISSUE), 175.929 cm3, 198.272 g 
      "Shape2":0 / "Shape2" / "Shape2"(G4Trd), 936 cm3, 1.85 g/cm3  (G4_BONE_COMPACT_ICRU), 936 cm3, 1.7316 kg
Calculating mass(es)...
Overall volume of "World":0, is 20736 cm3 and the daughter-included mass to unlimited depth is 12.8285 kg
G4ASCIITreeSceneHandler::EndModeling
\end{boxedverbatim}
}
\end{center}
\caption{Typical output of ASCII Tree.}
\label{figASCIITree}
\end{figure}

\section{Summary}

We have developed a visualisation system that is versatile, powerful and extensible.  It was designed to meet the needs of \Gfour users.  It supports several drivers over various graphics libraries, including OpenGL, Open Inventor and \Gfour's own tracking library ({\tt RayTracer}) and can write files for various browsers: DAWN, VRML, HepRApp.  It handles the \Gfour geometry hierarchy through a modeling library and can draw particle trajectories from stored events in a variety of ways.

\section*{Acknowledgments}

We wish to acknowledge the team of Geant4 collaborators who have made this work possible. In particular we wish to acknowledge the contribution of Satoshi Tanaka to the DAWN, VRML and gMocren drivers and to the DAWN and gMocren browsers. J.\,Allison wishes to acknowledged the support of The University of Manchester and Geant4 Associates International Ltd. L.\,Garnier wishes to acknowledge the support of Laboratoire de l'Acc\'{e}l\'{e}rateur Lin\'{e}aire, Orsay. This work is supported in part by the U.S. Department of Energy under contract number DE-AC02-76SF00515.


\begin{thebibliography}{99.}

\bibitem{Geant4}
S.\,Agostinelli et al, \Gfour, a simulation toolkit, Nuclear Instruments
and Methods in Physics Research, NIM A 506 (2003), 250-303.\\
See \Gfour Web page: \verb+http://cern.ch/geant4+.

\bibitem{DevsAndApps}
J.\,Allison et al., \Gfour Developments and Applications, IEEE Trans.\,Nucl.\,Sci.\,53, Issue: 1, Part 2 (2006) 270-278.

\bibitem{VisGen}
J.\,Allison et al., The \Gfour Visualisation System, Computer Physics Communications 178 (2008) 331-365.

\bibitem{UGT} The \Gfour User Guide for Toolkit Developers,
accessible from the \Gfour web page.\cite{Geant4}

\bibitem{SLACPresentation}
Introduction to \Gfour Visualization, Joseph Perl, Stanford Linear
Accelerator Center,\\
\verb+http://geant4.slac.stanford.edu/Presentations/vis/G4VisIntroduction.pdf+.

\bibitem{HepRep}
Joseph Perl, HepRep: a Generic Interface Definition for HEP Event
Display Representables,
Web page: \verb+http://www.slac.stanford.edu/~perl/heprep+

\bibitem{Qt}
The Qt Project, an open source cross-platform application and UI framework, \verb+http://qt-project.org/+.

\bibitem{UGA} The \Gfour User Guide for Application Developers,
accessible from the \Gfour web page.\cite{Geant4}  Of particular
interest and usefulness is Section 7.1, Built-in commands

\bibitem{bibOpenMotif}
For example, one may obtain Motif libraries from the Open Group,
\verb+http://www.opengroup.org/openmotif/+.

\bibitem{HepRApp} HepRApp: HepRep browser Application,\\
\verb+http:/www.slac.stanford.edu/~perl/HepRApp+.

\bibitem{VisCHEP97DAWN}
   S.\,Tanaka, M.\,Kawaguti, DAWN for \Gfour\ Visualization, 
   Proceedings of the CHEP '97 Conference, Berlin (Germany), April 1997.
   For documentation, see \verb+http://geant4.kek.jp/~tanaka/DAWN/About_DAWN.html+.

\bibitem{VisDAWNCUT}
   S.\,Tanaka, \verb+http://geant4.kek.jp/~tanaka/DAWN/About_DAWNCUT.html+.

\bibitem{VisDAVID}
   S.\,Tanaka, \verb+http://geant4.kek.jp/~tanaka/DAWN/About_DAVID.html+.

\bibitem{bibgMocren}
A.\,Kimura, S.\,Tanaka, K.\,Hasegawa, T.\,Sasaki,
``Visualization for Volume Data Scored by Geant4 Simulation",
IEEE Nuclear Science Symposium Conference Record, 
pp.2158-2161, October 2009. Also \verb+http://geant4.kek.jp/gMocren/+.

\end{thebibliography}
\end{document}